\documentstyle[amsbsy,epsf]{espart}

\newcommand{\C}{C_{\mathrm I,I'}} 
\newcommand{\Pro}[2]{\frac{\exp({\mathrm i}
    \mu |\vec{r}#1-\vec{r}#2|)} {|\vec{r}#1-\vec{r}#2|}}
\newcommand{\corl}[2]{\xi_{\mathrm #1,#2}} 
\newcommand{\pF}{p_{\mathrm F}}
\newcommand{\rp}{\rho_{\mathrm p}} 
\newcommand{\rn}{\rho_{\mathrm n}}
\newcommand{\tdensy}[2]{\rho^{(2)}_{\mathrm #1,#2}(\vec{r},\vec{r}')}
\newcommand{\im}{{\mathrm i}}
\newcommand{\tauk}{\tau_{\mathrm K}}

\newcounter{saveeqn}
\newcommand{\alpheqn}{\setcounter{saveeqn}{\value{equation}}\stepcounter{saveeqn}%
\setcounter{equation}{0}%
\renewcommand{\theequation}{%
              \mbox{\arabic{saveeqn}\alph{equation}}}}
\newcommand{\reseteqn}{\setcounter{equation}{\value{saveeqn}}%
\renewcommand{\theequation}{\arabic{equation}}}

\renewcommand{\thefootnote}{\fnsymbol{footnote}}

\def\bild#1#2{    
        \vspace*{-5mm}
        \begin{center}
        \begin{math}
        \epsfxsize#2cm
        \epsffile{#1}
        \end{math}
        \end{center}
        }

\begin{document}

\begin{frontmatter}
\title{Effective kaon mass in dense baryonic matter:\\ role of correlations}
\author{T. Waas \thanksref{gsi}}, \author{M. Rho\thanksref{sac}} and
\author{W. Weise \thanksref{gsi}}
\address{Physik-Department, Technische Universit\"at M\"unchen,\\ Institut f\"ur
Theoretische Physik, D-85747 Garching, Germany}
\thanks[gsi]{Work supported in part by GSI and BMBF}
\thanks[sac]{Permanent address: Service de Physique Th\'eorique, CEA Saclay,
F-91191 Gif-sur-Yette, France; supported in part by the A. v.~Humboldt Foundation}

\begin{abstract}
  We evaluate the effective kaon mass in dense nuclear matter. Pauli blocking
  and nucleon-nucleon short-range correlations are incorporated.  The effects
  of short-range correlations are shown to be moderate and figure importantly
  only at densities larger than 2 times normal nuclear density. We discuss the
  relations between the present results and the results obtained in
  next-to-next-to-leading order chiral perturbation theory (${{\cal O} (Q^3)}$,
  where $Q$ is the characteristic small energy-momentum scale probed). We also
  discuss mean-field aspects, with some remarks on the relation
  between the short-range correlations and the four-Fermi contact terms in the
  chiral effective Lagrangian.
\end{abstract}
\end{frontmatter}
\section{Introduction}
Much recent attention has been focused on the question how the mass of the
light pseudoscalar mesons (pions and kaons) change in nuclear matter with
increasing baryon density. In the case of pions, the answer was obvious for a
long time. In symmetric nuclear matter to leading order in the density
$\rho=\rho_{\mathrm p}+\rho_{\mathrm n}$ the pion mass shift is $\Delta
m^2_\pi=-4\pi b_0\rho$ where $b_0$ is the isospin-even $\pi N$ scattering
length.
A chiral low-energy theorem as well as $\pi N$ data imply that this $b_0$ is
very small. Effects of nuclear correlations are therefore important. A useful
and successful method to treat such correlations is the effective field
approach of Ericson and Ericson \cite{ericson}. The basic idea of this approach
is that repulsive two-body correlations between any pair of nucleons in the
medium reduce the pion field at the position of each scatterer. Consequently
$b_0$ is replaced by a (negative) effective scattering length $b_{\mathrm
eff}$. Its repulsive effect leads to a moderate upward shift of the
pion mass (by about 10\% at $\rho=\rho_0 =
0.17\mbox{ fm}^{-3}$, the density of normal nuclear matter).
This s-wave repulsion has been verified for many years in
pion-nuclear physics (see for example ref.~\cite{weise}). An impressive
confirmation has come very recently through the discovery of deeply bound
pion-nuclear states \cite{yamazaki}.

For kaons, the systematic exploration of in-medium masses has progressed
recently on the basis of the chiral SU(3) effective Lagrangian
\cite{brown,thomas2,thomas1,peter}. Here the driving s-wave interactions are
much stronger than those in the $\pi\mathrm N$ system. To leading order in the
kaon energy $\omega$ (or mass $m_{\mathrm K}$) the driving Born terms of the
threshold $\mathrm K^+N$ and $\mathrm K^-N$ scattering matrix\footnote[1]{The
scattering lengths $a$ are related to the threshold T-matrices by
$4\pi(1+m_{\mathrm K}/M)a=T$ where $M$ is the nucleon mass.} derived from the
chiral meson-baryon Lagrangian are \alpheqn
\begin{eqnarray}
  T_{\mathrm Born}({\mathrm K^-p})|_{\mathrm thr.}=-T_{\mathrm Born}({\mathrm
  K^+p})|_{\mathrm thr.} = \frac{m_{\mathrm K}}{f^2},\\
  T_{\mathrm Born}({\mathrm K^-n})|_{\mathrm thr.}=-T_{\mathrm Born}({\mathrm
  K^+n})|_{\mathrm thr.} = \frac{m_{\mathrm K}}{2f^2},
\end{eqnarray}
\reseteqn where $f\simeq 93$ MeV is the pseudoscalar decay constant.
These (Tomozawa-Weinberg, or TW) terms come from vector current
interactions. They are therefore repulsive for $\mathrm K^+N$ and attractive
for $\mathrm K^-N$ channels. Scalar interactions, attractive in both channels,
enter in next-to-leading order. They are proportional to the large kaon-nucleon
sigma term $\sigma_{\mathrm KN}$ of order $m^2_{\mathrm K}$, but compete with
repulsive $\omega^2$-terms of similar magnitude.

Based on these driving terms, in-medium calculations of kaon self energies have
been performed using either chiral perturbation theory \cite{brown}, or chiral
SU(3) dynamics \cite{thomas2,thomas1} in a non-perturbative coupled channels
scheme \cite{peter}. In the latter calculation the $\Lambda(1405)$ is generated
explicitly as a $\mathrm\overline{K}N$ ($\mathrm I=0$) quasibound state which quickly
dissolves, however, in nuclear matter and is thus of no relevance at higher
density. The primary effect in both types of calculations is a splitting of the
$\mathrm K^+$ and $\mathrm K^-$ masses in medium, a tendency which is indeed
already suggested by the TW Born terms (1).

The results of refs.~\cite{brown,thomas2,thomas1} include Pauli blocking which
turns out to be the dominant medium effect (corrections from nucleon binding
and Fermi motion are shown to be small \cite{thomas1}). On the other hand,
short-range nucleon-nucleon correlations were also suggested to be important
\cite{pethick} and should therefore be incorporated. The aim of the
present paper is to study effects of NN correlations on kaon masses together
with Pauli corrections using the effective field method of
ref.~\cite{ericson}. The formalism which is needed to adapt this method to the
kaon-nuclear case is presented in section 2. Since Pauli corrections have
already been evaluated \cite{thomas2} solving a coupled channel
Lippmann-Schwinger equation in medium, we can test the accuracy of the
effective field method at that level (section 3) and then continue to include
short-range NN correlations (section 4). In section 5 we discuss the relation
between this approach and four-Fermi contact terms of the chiral effective
Lagrangian used in chiral perturbation theory, and then close with concluding
remarks.

\section{The s-wave effective field for kaons in nuclear matter}
Here we apply the approach of ref.~\cite{ericson} to the s-wave kaon interaction
in nuclear matter. In spite of the similar nature of the kaon and the pion we
want to give a short review of the arguments which led Ericson and Ericson to
their famous optical potential. The reasons are the following:
\begin{enumerate}
\item The kaons have a different isospin structure.
\item The charge exchange channels were not included in their first work and
  only to second order in the later works (to our knowledge), whereas we need
  to iterate those channels to all orders for our purposes.
\item The Pauli principle plays an important role in ${\mathrm K}^-$-p
  scattering inside nuclear matter \cite{thomas1}. Therefore it has to be
  included to all orders in the optical potential.
\end{enumerate}

The starting point of ref.~\cite{ericson} is the observation that the
meson field $\phi(\vec{r})$ produced by an incident meson wave
$\exp(i\vec{k}\cdot\vec{r})$ on a system of (static) nucleons is the sum of
this incident wave and the scattered waves from all the individual 
nucleons. The scattered wave is simply the effective meson field
$\phi^{\mathrm{eff}}_i(\vec{r}_i)$ at the (fixed) scatterer $i$ times the scattering
amplitude ${\mathcal F}_i(\omega,\vec{k})$ times a spherical outgoing wave.
Thus,
\begin{equation} \label{phi}
  \phi(\vec{r})=\exp(\im\vec{k}\cdot\vec{r})+\sum_i \Pro{}{_i}\, {\mathcal F}_i
  (\omega,\vec{k})\,
  \phi^{\mathrm{eff}}_i(\vec{r}_i).
\end{equation}
Here $\omega$ and $\vec{k}$ denote the energy and momentum of the meson, and
$\mu=(\omega^2-m^2)^{1/2}$ with the meson rest mass $m$. We have
suppressed a phase $\exp(-\im \omega t)$, because we are only interested in
stationary solutions. 

Let us now identify the meson with a kaon. To include the charge-exchange 
channels of the kaon-nucleon system we recall that the scattering amplitude 
at scatterer $i$ has the following isospin structure:
\begin{equation}\label{isospin}
  { \mathcal F}_i(\omega,\vec{k})=f_{{\mathrm I}=0}(\omega,\vec{k})P_{{\mathrm
  I}=0}^i + f_{{\mathrm I}=1}(\omega,\vec{k})P_{{\mathrm
  I}=1}^i,
\end{equation}
where $P_{\mathrm I}^i$ projects on kaon-nucleon states of definite isospin I. 
With the (Pauli) isospin-matrices of the i'th scatterer (nucleon)
$\vec{\tau}_i$ and kaon $\vec{\tau}_{\mathrm K}$
they are given as:
\begin{equation}
  P_0^i=\frac{1-\vec{\tau}_{\mathrm K}\cdot\vec{\tau}_i}{4}, \quad
  P_1^i=\frac{3+\vec{\tau}_{\mathrm K}\cdot\vec{\tau}_i}{4}.
\end{equation}

In a homogeneous medium all effective fields $\phi^{\mathrm{eff}}_i(\vec{r})$
at scatterers $i$ with the same isospin quantum numbers are the same:
$P^i_{\mathrm{I}}\phi^{\mathrm{eff}}_i(\vec{r})$ can therefore be written as
$P^i_{\mathrm{I}}\phi^{\mathrm{eff}}_{\mathrm{I}}(\vec{r})$, dropping the
explicit reference to nucleon $i$.  We introduce the isospin densities
\begin{equation} \label{dens}
 \rho_{\mathrm{I}}(\vec{r})\equiv\langle0|
\sum_i P^i_{\mathrm{I}}\delta^3(\vec{r}-\vec{r}_i)|0\rangle 
\end{equation}
of the nuclear ground state $|0\rangle$ and rewrite eq.\ (\ref{phi}) as:
\begin{equation}\label{phi1}
  \phi(\vec{r})=\exp(\im\vec{k}\cdot\vec{r})+\sum_{\mathrm
  I=0,1}\int{\d}^3 r'\Pro{}{'}\,
  f_{\mathrm{I}}\,\rho_{\mathrm{I}}(\vec{r}')\phi^{\mathrm{eff}}_{\mathrm{I}} 
  (\vec{r}').
\end{equation}
The Klein-Gordon equation for $\phi(\vec{r})$ is then:
\begin{equation}
  \label{kg}
  \left(\omega^2+\nabla^2-m_{\mathrm K}^2\right)\phi(\vec{r})=-4\pi\sum_{\mathrm
  I=0,1}f_{\mathrm{I}}\,\rho_{\mathrm{I}}(\vec{r})
  \phi^{\mathrm{eff}}_{\mathrm{I}}
 (\vec{r}),
\end{equation}
where $m_{\mathrm K}$ is now the free kaon mass.

To solve this equation we need the effective field.  In complete analogy to
eq.\ (\ref{phi}), the effective field at a scatterer $i$ is given by the
incident wave and the scattered wave from all other particles with the
exception of $i$ itself:
\begin{equation} \label{phi_eff_c}
  \phi^{\mathrm{eff}}_i(\vec{r}_i)=\exp(\im\vec{k}\cdot\vec{r}_i)+
  \sum_{j\neq i} \Pro{_i}{_j}\, {\mathcal F}_j(\omega,\vec{k})\,
  \phi^{\mathrm{eff}}_j(\vec{r}_j).
\end{equation}
We now multiply eq. (\ref{phi_eff_c}) with $P_{\mathrm
  I}^i\delta^3(\vec{r}-\vec{r}_i)$ and sum over $i$. Using the isospin
structure of ${\mathcal F}_i$ as in eq.~(\ref{isospin}), taking the nuclear 
ground state expectation value and dividing by a common factor
$\rho_{\mathrm{I}}(\vec{r})$ one finds: 
\begin{eqnarray} \label{phi_eff_1}
  \phi^{\mathrm{eff}}_{\mathrm{I}}(\vec{r})&=&\exp(\im\vec{k}\cdot\vec{r})\\
 &+&\sum_{\mathrm{I}'=0,1}\int{\d}^3r'\Pro{}{'}\rho_{\mathrm{I}'}(\vec{r}') \left[1+
 \C(\vec{r},\vec{r}')\right]f_{\mathrm{I}'} \phi^{\mathrm{eff}}_{\mathrm{I}'}
 (\vec{r}') 
\nonumber\\
&=&\phi(\vec{r})+\sum_{\mathrm{I}'=0,1}\int{\d}^3r'\Pro{}{'}\rho_{\mathrm{I}'}
(\vec{r}') \C(\vec{r},\vec{r}')f_{\mathrm{I}'}\phi^{\mathrm{eff}}_{\mathrm{I}'}
(\vec{r}').\nonumber
\end{eqnarray}
Here we have introduced the NN pair correlation function
$\C(\vec{r},\vec{r}')$ which is related to the two-particle isospin density 
$\tdensy{I}{I'}$ as follows:
\begin{eqnarray}\label{twodensy}
  \tdensy{I}{I'}&\equiv&\langle0|\sum_i\sum_{j\neq i} P^i_{\mathrm{I}} P^j_{\mathrm{I}'}
 \delta^3(\vec{r}-\vec{r}_i)  \delta^3(\vec{r}'-\vec{r}_j)|0\rangle\nonumber\\
 &=&\rho_{\mathrm{I}}(\vec{r})\rho_{\mathrm{I}'} (\vec{r}')
  \left[1+\C(\vec{r},\vec{r}')\right].
\end{eqnarray}

Let us now consider the long-wavelength limit. In this limit, the pair
correlation function is non-vanishing only in a region small compared to the
kaon wavelength. The effective field at $\vec{r}'$ in eq.
(\ref{phi_eff_1}) can then be replaced by its value at $\vec{r}$. In the case
of nuclear matter the densities are constant and the pair correlation function
depends only on the distance $|\vec{r}-\vec{r}'|$ due to translation
invariance.  In this limit Eq. (\ref{phi_eff_1}) simplifies to:
\begin{equation} \label{couple}
  \phi^{\mathrm{eff}}_{\mathrm{I}}(\vec{r})
 =\phi(\vec{r})-\sum_{\mathrm{I}'=0,1}\corl{I}{I'}\rho_{\mathrm{I}'}
f_{\mathrm{I}'}\phi^{\mathrm{eff}}_{\mathrm{I}'}(\vec{r}) ,
\end{equation}
with constant densities $\rho_{\mathrm{I}'}$,  
where we have defined the correlation parameter $\corl{I}{I'}$ by:
\begin{equation}
  \label{cor_length}
  \corl{I}{I'}(\mu)=-\int{\d}^3 r'\Pro{}{'}\C(|\vec{r}-\vec{r}'|).
\end{equation}
When multiplied with density, this generalizes the isospin-independent inverse
correlation length $\langle 1/r\rangle$ of ref.~\cite{weise}.
Note that $\corl{I}{I'}$  is symmetric in $\mathrm I,I'$ and does not depend on
$\vec{r}$ because of translation invariance, so we can always set $\vec{r}=0$. Now the
two coupled equations (\ref{couple}) can easily be solved. Inserting the
effective fields $\phi^{\mathrm{eff}}_{\mathrm{I}}(\vec{r})$
into the wave equation (\ref{kg}) we obtain:
\begin{equation}
  \label{kg1}
  \left(\omega^2+\nabla^2-m^2_{\mathrm
  K}\right)\phi(\vec{r})=\Pi(\omega,\vec{k}) \phi(\vec{r}),
\end{equation}
with the self-energy, or optical potential,
\begin{eqnarray} \label{uopt}
  &&\Pi(\omega,\vec{k})=2\omega U_{\mathrm opt}(\omega,\vec{k})\\
  &&=-4\pi\frac{f_0\rho_0+f_1\rho_1-
  (2 \xi_{10} - \xi_{11} - \xi_{00})f_0f_1\rho_0\rho_1}
  {1 + f_0\xi_{00}\rho_0 + f_1\xi_{11}\rho_1 + (\xi_{00}\xi_{11} -
  \xi_{10}^2) f_0f_1\rho_0\rho_1}\nonumber.
\end{eqnarray}
This is the generalized s-wave optical potential of Ericson and Ericson. It includes
kaon-nucleon charge exchange and two-body NN correlations to all orders. It has
been derived here for infinitely heavy, static nucleons. In actual calculations
we multiply the usual kinematic factor $\sqrt{s}/M$ to each of the amplitudes
$f_{\mathrm I}$, where $\sqrt{s}$ is the kaon-nucleon center-of-mass energy and
$M$ the nucleon mass:
\begin{equation}
  \label{amplitudes}
  f_{\mathrm I}\rightarrow\frac{\sqrt{s}}{M}f_{\mathrm I}.
\end{equation}

The effective kaon mass $m^\star_{\mathrm K}$ is defined by the energy $\omega$ of a kaon at
rest in matter ($\vec{k}=0$, $\phi(\vec{r})=\mathrm const.$). The corresponding
(off-shell) s-wave scattering amplitudes are expressed in terms of the energy
dependent (off-shell) scattering lengths $a(\omega)$. Their threshold values
$a(\omega=m_{\mathrm K})$ coincide with the physical scattering lengths. In the
sector with strangeness ${\mathrm S}=-1$:
\begin{eqnarray}
  f_0^{\mathrm S=-1}(\omega,\vec{k}=0)&=&2\,a({\mathrm
K^-p\rightarrow K^-p})-a({\mathrm K^-n\rightarrow K^-n}),\nonumber\\
  f_1^{\mathrm S=-1}(\omega,\vec{k}=0)&=&a({\mathrm K^-n\rightarrow K^-n}).
\end{eqnarray}
Analogous relations hold in the ${\mathrm S}=+1$ sector with the $\mathrm{K}^+N$
scattering lengths:
\begin{eqnarray}
  f_0^{\mathrm S=+1}(\omega,\vec{k}=0)&=&2\,a({\mathrm
K^+n\rightarrow K^+n})-a({\mathrm K^+p\rightarrow K^+p}),\nonumber\\
  f_1^{\mathrm S=+1}(\omega,\vec{k}=0)&=&a({\mathrm K^+p\rightarrow K^+p}).
\end{eqnarray}

In order to calculate the off-shell scattering lengths we use the coupled
channels approach of ref.~\cite{peter} which is based on chiral dynamics and
describes all available low-energy kaon-nucleon data together with
photoproduction channels.

Note that the energy of a $\mathrm K^-$ in matter is generally complex, its
imaginary part being related to processes in which $\mathrm\overline{K}N$
decays into pion-hyperon channels. But as we will see, this imaginary part is always small compared with the real part. Therefore,
the effective kaon mass ($m_{\mathrm K}^\star\equiv\mathop{\mathrm Re}\omega$) is determined to
a good approximation through the simpler relation:
\begin{equation} \label{real_disp}
  (m_{\mathrm K}^\star)^2-m_{\mathrm K}^2=2m_{\mathrm K}^\star\,\mathop{\mathrm Re}U_{\mathrm opt}(m^\star_{\mathrm K},\vec{k}=0).
\end{equation}
The effective kaon decay width
$\Gamma=-2\mathop{\mathrm{Im}}\omega$, can then be approximated by:
\begin{equation} \label{imag_disp}
  \Gamma=-2\mathop{\mathrm Im}U_{\mathrm opt}(m^\star_{\mathrm K},\vec{k}=0).
\end{equation}

\section{Pauli principle effects}
In this section we calculate the average inverse correlation length for a
Fermi-gas. In this model the nucleons do not interact and the correlation
length is entirely due to the Pauli exclusion principle. The nuclear ground
state $|0\rangle$ is a Slater determinant of plane-waves. All proton
(neutron) levels are filled up to the Fermi momentum $\pF^{\mathrm p}$
($\pF^{\mathrm n}$).  The nucleon states are characterized as usual by their momentum
$\vec{p}$, spin $s=\pm\half$ and isospin $t=\pm\half$ for (p,n). In the
following $\alpha$ and $\beta$ denote occupied states.

The isospin densities $\rho_{\mathrm I}$ defined in eq.\ (\ref{dens}) are given
by:
\begin{eqnarray}
  \rho_0&=&\frac{1}{4}\left(\rp+\rn\right)-\frac{\tauk^3}{4}\left(\rp-\rn\right)\\
  \rho_1&=&\frac{3}{4}\left(\rp+\rn\right)+\frac{\tauk^3}{4}\left(\rp-\rn\right),
  \nonumber
\end{eqnarray}
where $\rp=(\pF^{\mathrm p})^3/3\pi^2$ and $\rn=(\pF^{\mathrm n})^3/3\pi^2$ are
the proton and neutron densities, and $\tauk^3=\pm 1$ for $\mathrm K^+$ or
$\mathrm K^-$, respectively.

In the Fermi gas model the two-particle isospin density (\ref{twodensy}) becomes
\begin{eqnarray}\label{sec}
  \tdensy{I}{I'} &=&\sum_{\alpha\beta}\langle\alpha\beta|P_{\mathrm I}^1P_{\mathrm
  I'}^2 \delta^3(\vec{r}-\vec{r}_1)
  \delta^3(\vec{r}'-\vec{r}_2)|\alpha\beta-\beta\alpha\rangle\nonumber\\
&=&\sum_\alpha\langle\alpha|P_{\mathrm I}\delta^3(\vec{r}-\vec{r}_1)|\alpha\rangle 
\sum_\beta\langle\beta|P_{\mathrm I'}\delta^3(\vec{r}'-\vec{r}_2)|\beta\rangle
\nonumber\\
&-&\sum_{\alpha,\beta}\langle\alpha|P_{\mathrm I}\delta^3(\vec{r}-\vec{r}_1)
|\beta\rangle\langle\beta|P_{\mathrm
I'}\delta^3(\vec{r}'-\vec{r}_2)|\alpha\rangle
\nonumber\\
&=&\rho_{\mathrm I}(\vec{r})\,\rho_{\mathrm I'}(\vec{r}')\left[1+\C^{\mathrm
Pauli}(|\vec{r}-\vec{r}'|)\right]
\end{eqnarray}
where the single particle matrix elements are understood to involve integrals
over $\vec{r}_1,\vec{r}_2$. The result for the Pauli correlation function is:
\begin{equation} \label{cori}
  \C^{\mathrm Pauli}(r)=-\half\sum_{t,t'}
  c_t(r)c_{t'}(r)\frac{\langle t|P_{\mathrm I}|t'\rangle\langle t'|P_{\mathrm
  I'}|t\rangle}{\rho_{\mathrm I}\,\rho_{\mathrm I'}}
\end{equation}
where
\begin{equation}
  c_t(r)=\frac{3j_1(\pF^t r)}{\pF^t r}\rho_t
\end{equation}
with the spherical Bessel function $j_1$, and the indices $t$, $t'$ distinguish
between protons and neutrons. 

Inserting this into eq.\ (\ref{cor_length}) we determine the correlation
parameters $\corl{I}{I'}$ for the following cases:

i) {\bf $\mathbf K^{\bf +}$ or $\mathbf K^{\bf -}$ in symmetric nuclear matter}

\nopagebreak[4]

Here we have $\pF^{\mathrm p}=\pF^{\mathrm n}=\pF$, $\rp+\rn=\rho$ and
$\rho_0=\rho_1/3=\rho/4$. One finds
\begin{equation}
  \xi\equiv\xi_{00}=3\,\xi_{11}=\frac{9}{\pF^2}F(0,\mu,\pF),
\end{equation}
\begin{displaymath}
  \xi_{10}=\xi_{01}=0,
\end{displaymath}
where we have introduced
\begin{equation}
  F(R,\mu,\pF)=4\pi\int_R^\infty{\d} r\frac{\exp({\mathrm i} \mu r)}{r}j^2_1(\pF r) 
\end{equation}
and recall that $\mu(\omega)=\sqrt{\omega^2-m^2_{\mathrm K}}$. The optical potential
(\ref{uopt}) for this case becomes
\begin{equation}
  2\omega U_{\mathrm opt} = -4\pi\left[\frac{1}{4}\,\frac{f_0\rho}{1
  +\frac{\xi}{4}f_0\rho}+\frac{3}{4}\,\frac{f_1\rho}{1
  +\frac{\xi}{4}f_1\rho} 
  \right].
\end{equation}
To leading order in $\xi$ one finds
\begin{equation}
  2\omega U_{\mathrm opt} = -4\pi{\mathcal F}_{\mathrm eff}\rho
\end{equation}
with the effective amplitude (including the $\sqrt{s}/M$ factors):
\begin{equation}
  {\mathcal F}_{\mathrm eff}=\frac{\sqrt{s}}{4M}(f_0+3f_1) - \frac{s}{16M^2}
  (f_0^2+3f_1^2) \xi\rho+{\mathcal O}(\xi^2).
\end{equation}
At threshold $(\omega=m_{\mathrm K})$ we have $\sqrt{s}/M=1+m_{\mathrm K}/M$,
$\mu=0$, and $\xi\rho$ reduces to the familiar inverse correlation length
\cite{ericson,weise}:
\begin{equation}
  [\xi_{00}\rho_0]^{\mu=0}_{\omega=m_{\mathrm K}} = \langle\frac{1}{r}\rangle = -\int{\d}^3r\frac{C_{00}^{\mathrm Pauli}(r)}{r}\rho_0=\frac{3\pF}{2\pi}.
\end{equation}

ii) {\bf $\mathbf K^{\bf +}$ in neutron matter}

Here we have $\pF^{\mathrm p}=0$, $\pF^{\mathrm n}=\pF$ and
$\rho_0=\rho_1=\rn/2$.
One finds
\begin{equation}
  \xi\equiv\xi_{00}=\xi_{11}=\xi_{10}=\xi_{01}=\frac{9}{2\pF^2}F(0,\mu,\pF).
\end{equation}
The $\mathrm K^+$ self-energy in neutron matter becomes
\begin{equation}\label{pin}
  \Pi_{\mathrm n}=-4\pi\frac{{\mathcal F}\rho_{\mathrm n}}{1+{\mathcal
  F}\xi\rho_{\mathrm n}} 
\end{equation}
with
\begin{equation}
  {\mathcal F}=\frac{\sqrt{s}}{2M}(f_0+f_1)
\end{equation}

iii) {\bf $\mathbf K^{\bf -}$ in neutron matter}

In this case only the $I=1$ amplitude contributes and we have (with
$\pF^{\mathrm n}=\pF$, $\rho_0=0$, $\rho_1=\rn$):
\begin{equation}
  \xi\equiv\xi_{11}=\frac{9}{2\pF^2}F(0,\mu,\pF).
\end{equation}

The $\mathrm K^-$ self-energy in neutron matter has the same form as
$\Pi_{\mathrm n}$ of eq.~(\ref{pin}), but now with
\begin{equation}
  {\mathcal F}=\frac{\sqrt{s}}{M}f_1.
\end{equation}
This completes our exposition of Pauli exchange effects on the kaon
self-energies. Using those self-energies we have solved the dispersion
relations for the effective kaon masses (and decay widths) in matter. The
results for $\mathrm K^+$ and $\mathrm K^-$ in symmetric nuclear matter are
shown in Fig.~1, those for neutron matter in Fig.~2. Turning on the Pauli
correlations moves the $\mathrm K^+$ and $\mathrm K^-$ masses upward from the
values they would have with just the leading order self-energies
$\Pi=-4\pi(\sqrt{s}/M)[f_0\rho_0+f_1\rho_1]$. Note that with the TW Born terms
(1) alone we would have
\begin{equation}
  \Pi^\pm_{\mathrm Born}=\pm
  \frac{\omega}{f^2}\left[\frac{3}{4}(\rp +
  \rn)+\frac{1}{4}(\rp-\rn)\right] 
\end{equation}
for $\mathrm K^+$ or $\mathrm K^-$, respectively. This already accounts for the
qualitative feature of $\mathrm K^+/K^-$ mass splitting in matter. Iterations
of the Born amplitudes as in \cite{thomas2} reduce the repulsive $\mathrm K^+N$
and enhance the attractive $\mathrm K^-N$ interactions, leading to the patterns
seen in Figs.~1 and 2. The Pauli correlations act repulsive on both the
$\mathrm K^+$ and $\mathrm K^-$ branches.

In Fig.~1 and 2 we also compare the present treatment of Pauli correlations to
the results of ref.~\cite{thomas2} where these effects have been treated by
solving an in-medium coupled channels Lippmann-Schwinger equation. The
agreement between the present effective field approach and the previous full
coupled-channels result is quite remarkable, especially in view of the fact
that the Lippmann-Schwinger coupled channels calculation does {\em not\/} make
use of the fixed scatterer approximation.

\section{Short-range correlations}
The Pauli exchange correlations discussed in the previous section give an
impression of the long-range part of the nucleon-nucleon pair correlation
function. As a next step we study the influence of short-range repulsive NN
correlations. To start with, consider a schematic hard core.
Assume that the two-particle density (\ref{twodensy}) vanishes at distances
smaller than $R$ and that the pair correlation
function is therefore simply
\begin{equation}\label{shortrange}
  \C(r)=-1 \mbox{ for $r\le R$}
\end{equation}
independent of spin and isospin.

It is instructive to compare this with the Pauli pair correlation functions
(\ref{cori}) at zero distance ($\vec{r}=\vec{r}'$). For example, in the
symmetric nuclear matter case they are:
\begin{eqnarray}
  C^{\mathrm Pauli}_{00}(0)= -1,\quad C^{\mathrm Pauli}_{10}(0)=C^{\mathrm
  Pauli}_{01}(0)=0,\quad C^{\mathrm Pauli}_{11}(0)=-\frac{1}{3}. 
\end{eqnarray}
Therefore the main contributions of the short-range correlations to the pair
correlation functions are in the sectors ${\mathrm (I,I')}=(0,1),(1,0),(1,1)$.
In the sector ${\mathrm (I,I')}=(0,0)$ the Pauli principle already prevents the
two nucleons to come close to each other, so the repulsive core is nearly
inactive.

For distances larger than $R$ we assume that the pair correlation function is
completely described by Pauli exchange effects. In this simple model the
correlation parameter $\corl{I}{I'}$ is given by:
\alpheqn
\begin{equation}\label{corl2}
 \corl{I}{I'}(R,\omega,\pF)=G(R,\mu,\pF)+9\pF^{-2}({\mathrm
  1-I-I'+\frac{4}{3}II'})F(R,\mu,\pF)
\end{equation}
for symmetric nuclear matter and by
\begin{equation}\label{corl3}
 \corl{I}{I'}(R,\omega,\pF)=G(R,\mu,\pF)+\frac{9}{2}\pF^{-2}F(R,\mu,\pF)
\end{equation}
\reseteqn
for $\mathrm K^+$ in neutron matter (for the $\mathrm K^-$ in neutron matter,
multiply $F$ by $\delta_{\mathrm I1}\delta_{\mathrm I'1}$).
Here the function
\begin{equation}
  G(R,\mu,\pF)=4\pi\int_0^R{{\d} r}\, r \exp(\im \mu r)=
  \frac{4\pi}{\mu^2}\left[\left(1-\im \mu R\right)\exp(\im \mu R)-1\right]
\end{equation}
describes the short-range correlations (\ref{shortrange}).  
Later we will discuss a different parameterization of the short-range part, one
which approximates the realistic G-matrix derived from Brueckner Hartree Fock
calculations. 

Inserting the correlation parameters (\ref{corl2},b) into the optical potential
(\ref{uopt}), the resulting solutions of the kaon dispersion relations
(\ref{real_disp},\ref{imag_disp}) are shown in Fig.~3 (symmetric nuclear
matter) and in Fig.~4 (neutron matter). For comparison we show
in these Figures also the effective kaon masses calculated in the previous
section with Pauli blocking only ($R=0$). The effect of the
short-range correlations starts
out small at low densities, becoming sizeable only at densities larger than  
normal nuclear matter density, $\rho_0=0.17\,\mbox{fm}^3$.  The short-range
correlations increase both kaon and antikaon masses as compared with the
results including Pauli blocking only. 

A better parameterization of the pair
correlation function in symmetric nuclear matter is given in \cite{wolfram}. 
There the short-range part is chosen to approximate the Brueckner G-Matrix and
found to be quite well
described by a spherical Bessel function:
\begin{equation}\label{g-matrix}
  \C^{\mathrm SR}(\omega,r)=-j_0(q_c r);\quad  q_c=3.93 \mbox{ fm}^{-1}
\approx m_{\omega},
\end{equation}
where $m_{\omega}$ is the mass of the $\omega$-meson.
To incorporate the Pauli part of the pair correlation function (\ref{cori}) we
assume that these two correlations contribute multiplicatively in 
the two-particle density:
\begin{equation}
   \tdensy{I}{I'}=\rho_{\mathrm I}(\vec{r}) \rho_{\mathrm I'}(\vec{r}') \left[
   1 + \C^{\mathrm Pauli}(|\vec{r}-\vec{r}'|) \right]\left[1 + \C^{\mathrm SR}(|\vec{r}-\vec{r}'|)
   \right].
\end{equation}
Therefore the full pair correlation function is
\begin{equation}
  \C(r)=\C^{\mathrm Pauli}(r)+\C^{\mathrm SR}(r)+\C^{\mathrm Pauli}(r)\C^{\mathrm SR}
\end{equation}
in this model. The effective kaon masses, calculated with these pair
correlation functions, are 
also shown in Fig.~3 and differ only marginally from the ones
calculated with the simpler ansatz (\ref{shortrange}).

While the repulsive effect of short range correlations is moderate, we expect
that it will be important for the issue of kaon condensation in neutron stars 
unless certain
nonperturbative effects discussed below (which are not included in the present
treatment) bring large corrections. Kaon
condensation can occur once the effective kaon mass gets equal to or smaller 
than the electron
chemical potential \cite{brown}. Without the short-range correlations this
matching will occur around $\rho\approx3\,\rho_0$ if one uses 
the electron chemical potential employed in \cite{brown}. 
With the inclusion of the short-range correlations this
matching will be shifted to higher density $\rho>3\rho_0$ (see Fig.~5) or if
hyperons are present in  
the dense matter as suggested in some recent papers
\cite{schaffner,prakash}, may not even take place.
In the latter case, the electron
chemical potential can drop at densities
$\rho${\raisebox{-0.8ex}{$\stackrel{\textstyle >}{\sim}$}}$ 3\,\rho_0$ due to the appearance of
$\Sigma^-$-- and $\Xi^-$--hyperons in the neutron star matter. This matter
is currently under debate and deserves to be clarified.

\section{Comparison with higher-order chiral perturbation 
and mean-field calculations}
We shall now compare
the present calculation which, as we should stress, is very tightly constrained
\cite{peter} by the full ensemble of kaon-nucleon data, to the chiral
perturbation calculation 
of Lee et al. \cite{LBMR} and the mean-field treatment of \cite{BR96}, both
of which are less constrained by on-shell data.
Some features between them overlap and some do not. 
Understanding the similarities and differences between these approaches
will be crucial in confronting data on the properties of kaons in dense
matter coming from extrapolations of kaonic atom data as well as the GSI
experiments of the KaoS and FOPI collaborations currently in progress. 
It will also be important for establishing whether or not kaons can
condense at low enough density as to be relevant to ``nuclear star"
matter \cite{brownbethe}.

To streamline the discussion, we first
recall the key ingredients of the present
treatment and of ref.~\cite{LBMR}. In the present paper (call it A), 
the potential for 
kaon-nuclear interactions is constructed with ``irreducible graphs" calculated
to ${{\cal O} (Q^2)}$  in chiral perturbation theory 
and then inserted
as the kernel into the Lippmann-Schwinger equation for both free-space and 
in-medium kaon-nucleon interactions. The solution of the integral equation
correctly accounts for the
infrared singular ``reducible" graphs to {\em all\/} orders of the chiral 
expansion in a way consistent with the general strategy of 
chiral perturbation
theory in the presence of bound states or resonances. Pauli correlations are
readily taken into account at the level of solving the Lippmann-Schwinger 
equation 
or as, in this paper, in the effective field method. While these 
two approaches
are found to give the same result,  the latter has the advantage
that the short-range 
correlations can also be simply implemented.

In ref.~\cite{LBMR}, the in-medium kaon-nuclear interaction is treated to  
one-loop order (or ${{\cal O} (Q^3)}$) with the quasi-bound $\Lambda (1405)$
introduced as a local interpolating field. 
While the $\Lambda (1405)$ as an elementary local field
may not be accurate enough
for ${\mathrm K}^-$-proton scattering 
near threshold, it can however be an adequate way of 
introducing the $\Lambda (1405)$ degree of freedom
for in-medium physics (a discussion on this point is found 
in \cite{LMR96} where 
a chiral expansion to ${{\cal O} (Q^2)}$ is made. 
To be more predictive, one would 
have to do an ${{\cal O} (Q^3)}$ calculation which is not feasible at the
moment). To the chiral order considered, the
$\Lambda (1405)$ can contribute in two ways: one, in the off-shell elementary
kaon-nucleon scattering amplitude
and two, in an effective four-Fermi interaction 
involving both nucleon and $\Lambda (1405)$. In this approach (that we shall
refer to as B), a certain class of two-body 
correlations including Pauli-principle effects
are included but the short-range correlations calculated in A 
are not.

Let us now compare quantitatively the two (A and B) approaches. For in-medium 
${\mathrm K}^+$ properties, the results are quite similar: 
the effective kaon mass $m_K^\star$ scales in density at  
about the same rate up to the normal matter density and increases somewhat
more rapidly in A than in B. The reason for this similarity
is simple:  since 
the $\Lambda (1405)$ does not figure importantly in this channel, its 
ramifications in A (i.e., the proper treatment of the binding mechanism of 
the $\Lambda (1405)$) and in B (i.e., the four-Fermi interaction involving 
$\Lambda (1405)$) are unimportant.

The situation with ${\mathrm K}^-$-nuclear interaction is, however, 
considerably different. This is more pronounced in symmetric nuclear matter 
than neutron matter. 
The approach A in symmetric nuclear matter predicts a 
decrease in the ${\mathrm K}^-$ 
mass (or equivalently the attraction in the kaon-nuclear potential) of 
about 120 MeV whereas in
the approach B, the attraction is sizeable greater, say,
about 200 MeV. This large attraction in
B is obtained 
by fixing by fiat a parameter associated with four-Fermi interactions
involving the $\Lambda (1405)$ to agree with the 
attraction suggested by Friedman, Gal and Batty \cite{gal}
from their analysis of kaonic atoms. It is {\it not} a 
prediction of the theory. 
The resulting attraction at densities greater than that of nuclear matter
found in B is thus simply explained by this additional term.
The four-Fermi interaction in question corresponds to an 
${{\cal O} (Q^3)}$ irreducible graph in the chiral expansion and is not 
included in the approach A. One can think of the short-range 
correlations calculated in A as a four-Fermi interaction 
involving nucleons only, but in a naive chiral counting it would appear at 
${{\cal O} (Q^5)}$. Since in B {\em not\/} all the
four-Fermi interaction terms that can contribute are included, the additional
attraction in B over and above
that accounted in A can be thought 
of as the net effect of the irreducible ${{\cal O} (Q^3)}$ terms not taken into
account in A. That it is an additional attraction is a consequence of the
fitting.
As such, it will be subject to further empirical constraints.

The mean-field approach of \cite{BR96} exploits the fact that
in medium, the $\Lambda (1405)$ plays an insignificant role (due to the 
dissociation of the $\Lambda (1405)$ 
in the approach A or due to a weak coupling of the $\Lambda (1405)$ with no
kinematic enhancement near threshold in the approach B). 
As such, this approach is not constrained by the on-shell kaon-nucleon
data as in the cases of A and B.  Limiting to the
${{\cal O} (Q^2)}$ chiral order and ignoring the role of the $\Lambda (1405)$, 
all dominant in-medium higher-order correlation effects 
(particularly, the many-Fermi
interactions in the scalar channel) are then subsumed into the scaling of the
effective pseudoscalar decay constant $f_\pi^\star$.\footnote{This can be understood as follows. 
N-Fermi interactions  increase the chiral order as 
${Q^{N/2}}$. Thus higher-N Fermi interactions are generally 
suppressed. There is, however, {\it one possible exception}: 
when the interaction
is mediated by a scalar meson, then the corresponding N-Fermi interactions can
give a coherent effect when the effective mass of the scalar meson 
becomes light as it is expected to 
happen in dense matter (or hot matter near the chiral
phase transition). Such a scalar-induced higher-Fermi interaction then
acts like a scalar tadpole on hadrons and can give rise to a decreasing
hadron mass or equivalently to the $f_\pi^\star$ \cite{BRscaling}.}
This leads to an effective constituent quark -- or mean-field -- counting
that predicts that the effective scalar and vector potentials in kaon-nuclear
interactions 
-- the kaon containing only one valence chiral quark --
are 1/3
of the corresponding nucleon-nuclear interactions that involve three light
quarks. This description which correctly accounts for the nucleon
mean field in nuclear matter provides a simple explanation for
the large ($\sim 200$ MeV) attraction for kaonic atoms of ref.\cite{gal}
and 
seems to be consistent with the (preliminary) KaoS and FOPI data of
kaon properties in dense nuclear medium \cite{gerry}.

Whether or not such an
additional higher order chiral contribution with a net attraction
is unambiguously needed
for understanding in-medium kaon properties
should be 
settled by careful analyses of the ongoing experiments on kaons in 
dense medium.  These experiments will also provide a crucial constraint on
the role of kaons in neutron-rich medium and determine in particular
whether kaon condensation does actually figure in the structure of 
compact stars.
If such an attraction is established, it would help in going beyond
the present treatment in a systematic chiral perturbation calculation 
that can account {\em simultaneously\/}
for kaon-nucleon and kaon-nuclear interactions.   
\newpage

{\bf Acknowledgements}

One of the authors (MR) acknowledges the support of the Humboldt Foundation
through its ``Forschungspreis" and the generous 
hospitality of the Institut f\"ur Theoretische Physik, Technische
Universit\"at M\"unchen where this work was done. He is grateful for 
illuminating comments from Gerry Brown and C.-H. Lee.

One of us (WW) gratefully acknowledges support as a recipient of a
Humboldt-Mutis award. He thanks Enlogio Oset for his kind hospitality and
discussions at the University of Valencia where this paper was completed.

Two of us (TW and WW) thank Avraham Gal for helpful discussions and comments.

\newpage

\centerline{\bf Figure Captions}

\bigskip

{\bf Fig.~1:} Effective $\mathrm K^-$ and $\mathrm K^+$ masses in
symmetric nuclear matter as a function of the density $\rho=\rho_{\mathrm
p}+\rho_{\mathrm n}$, in units of nuclear matter density $\rho_0=0.17
\mbox{ fm}^{-3}$. The full curves are the results of this
work, including only Pauli correlations. The dashed lines are the results of
the in-medium coupled channels calculation of ref.~\cite{thomas2}. The dotted
lines show results without Pauli correlations. The
dashed--dotted lines show for reference the effective masses obtained with the
Tomozawa-Weinberg Born terms (1) only.

\bigskip
{\bf Fig.~2:} Effective $\mathrm K^-$ and $\mathrm K^+$ masses in
neutron matter as a function of the neutron density $\rn$. Notations as in Fig.~1.

\bigskip

{\bf Fig.~3:} Effective $\mathrm K^-$ and $\mathrm K^+$ masses in
symmetric nuclear matter as a function of density $\rho$. Dashed lines are the
results including only Pauli blocking, 
full lines include also the short-range correlation functions (\ref{shortrange}). The
dashed--dotted lines are calculated with the short-range correlation function
(\ref{g-matrix}) for comparison. The lower section shows the $\mathrm K^-
$ width in matter (scaled by a factor 100).   

\bigskip
{\bf Fig.~4:} Effective $\mathrm K^-$ and $\mathrm K^+$ masses in
neutron matter as a function of the neutron density. Dashed lines
are the results including only Pauli blocking and the full curves include also
short-range correlations as in eq.~(\ref{shortrange}). The lower section shows
the $\mathrm K^-$ width in matter (scaled by a factor 100).

\bigskip
{\bf Fig.~5:} The effective $\mathrm K^-$ mass in neutron star matter 
(full line: with short range
correlations; dashed line: without short range correlations) versus the electron
chemical potential (dotted line). For demonstration we have used the chemical
potential from ref.~\cite{LMR96} calculated with one of their parametrizations
of the symmetry energy (case ``$F(u)=u$").

\newpage
\vspace*{-1cm}
\bild{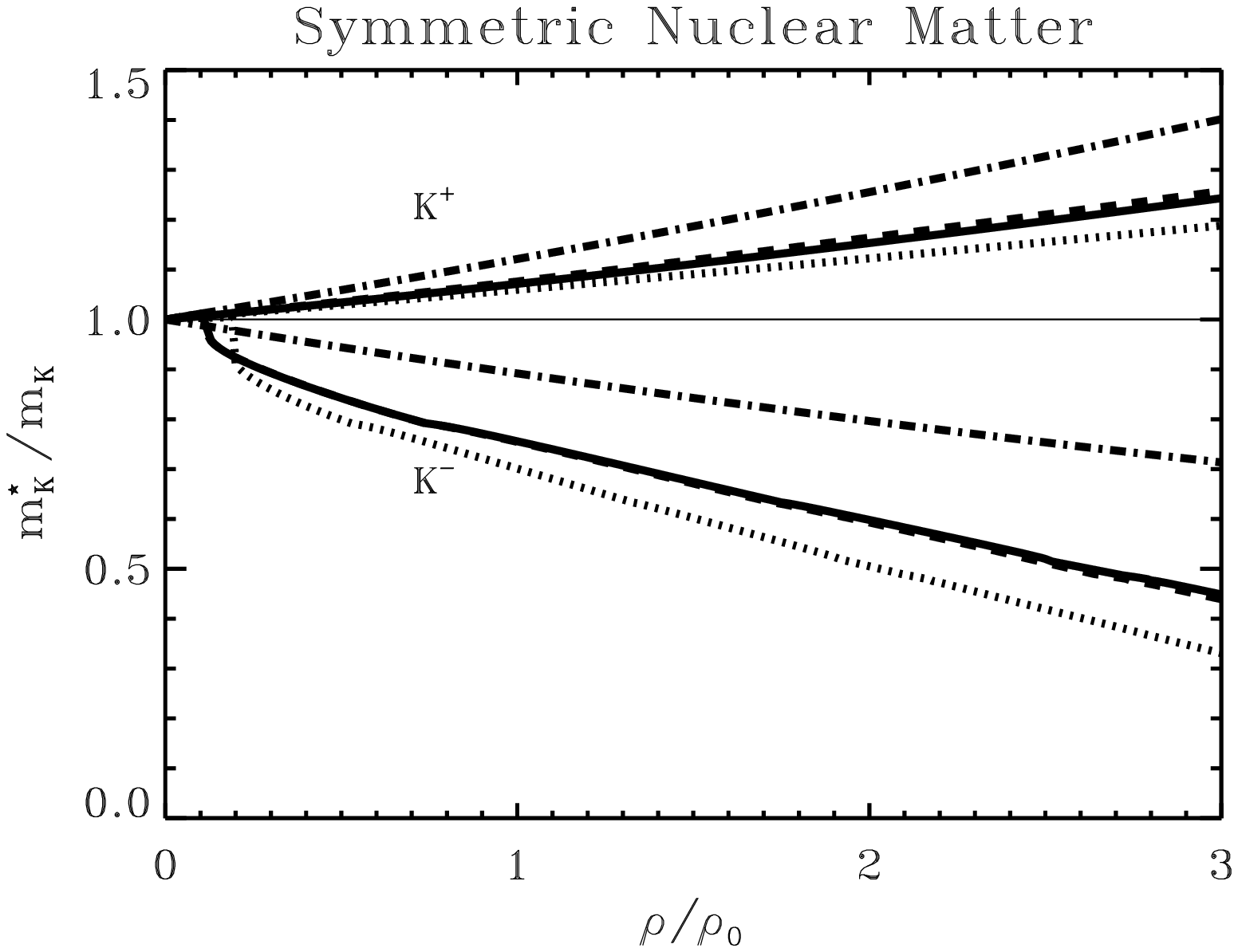}{13.8}
\begin{center}
\vspace*{-.5cm}
\large Figure 1
\end{center}

\vspace*{1.2cm}

\bild{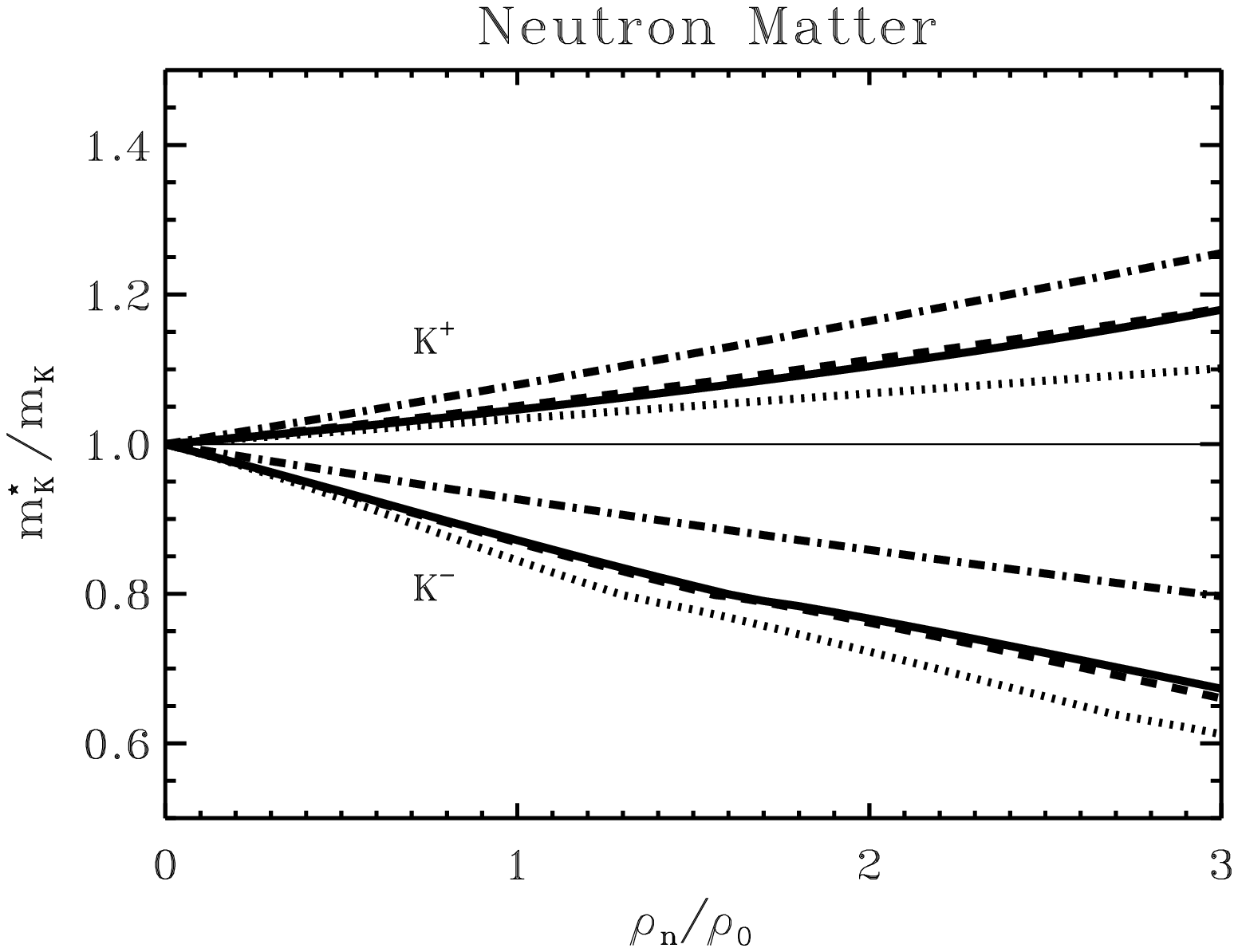}{13.8}
\begin{center}
\vspace*{-.5cm}
\large Figure 2
\end{center}

\newpage
\vspace*{-1cm}
\bild{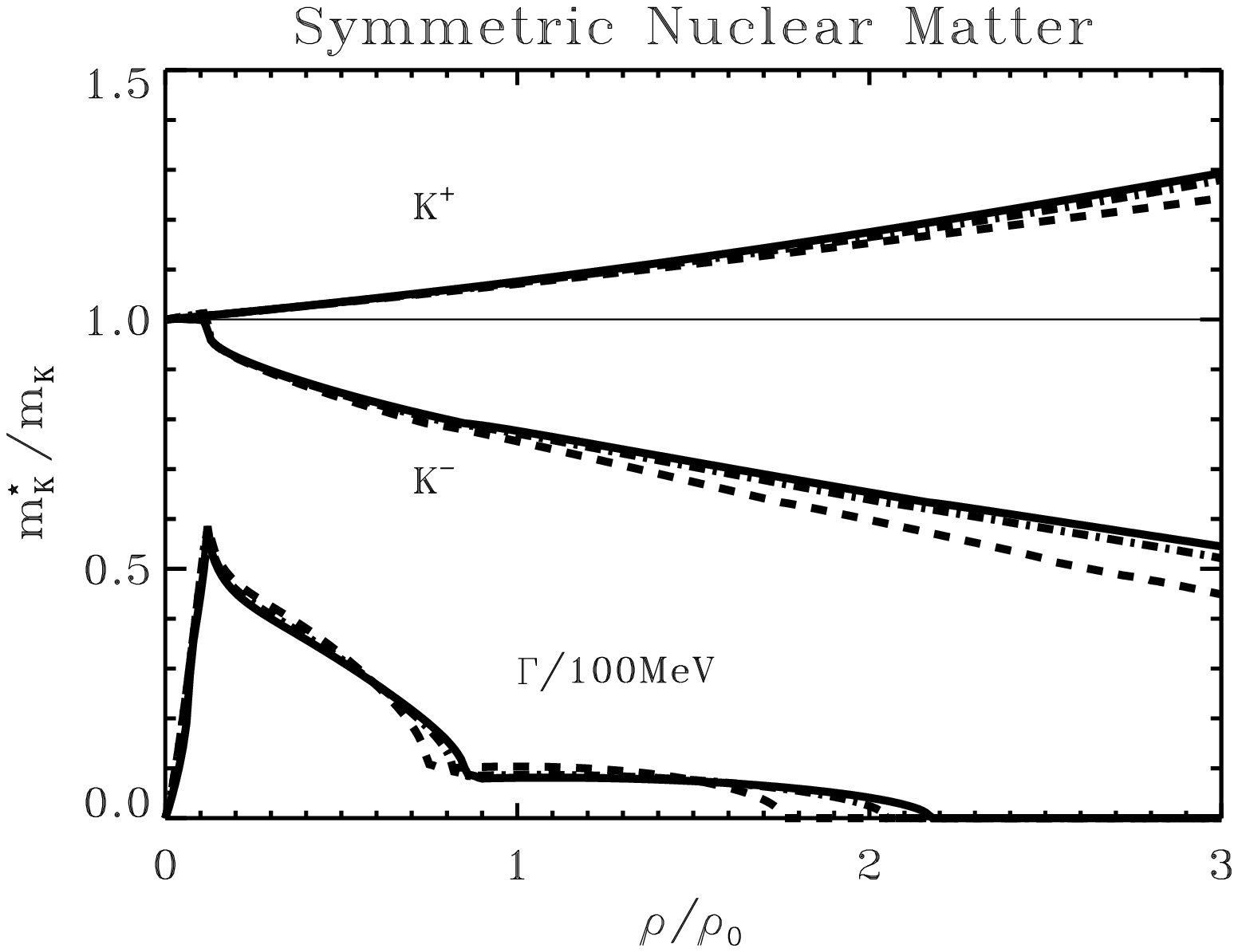}{13.8}
\begin{center}
\vspace*{-.5cm}
\large Figure 3
\end{center}

\vspace*{1.2cm}

\bild{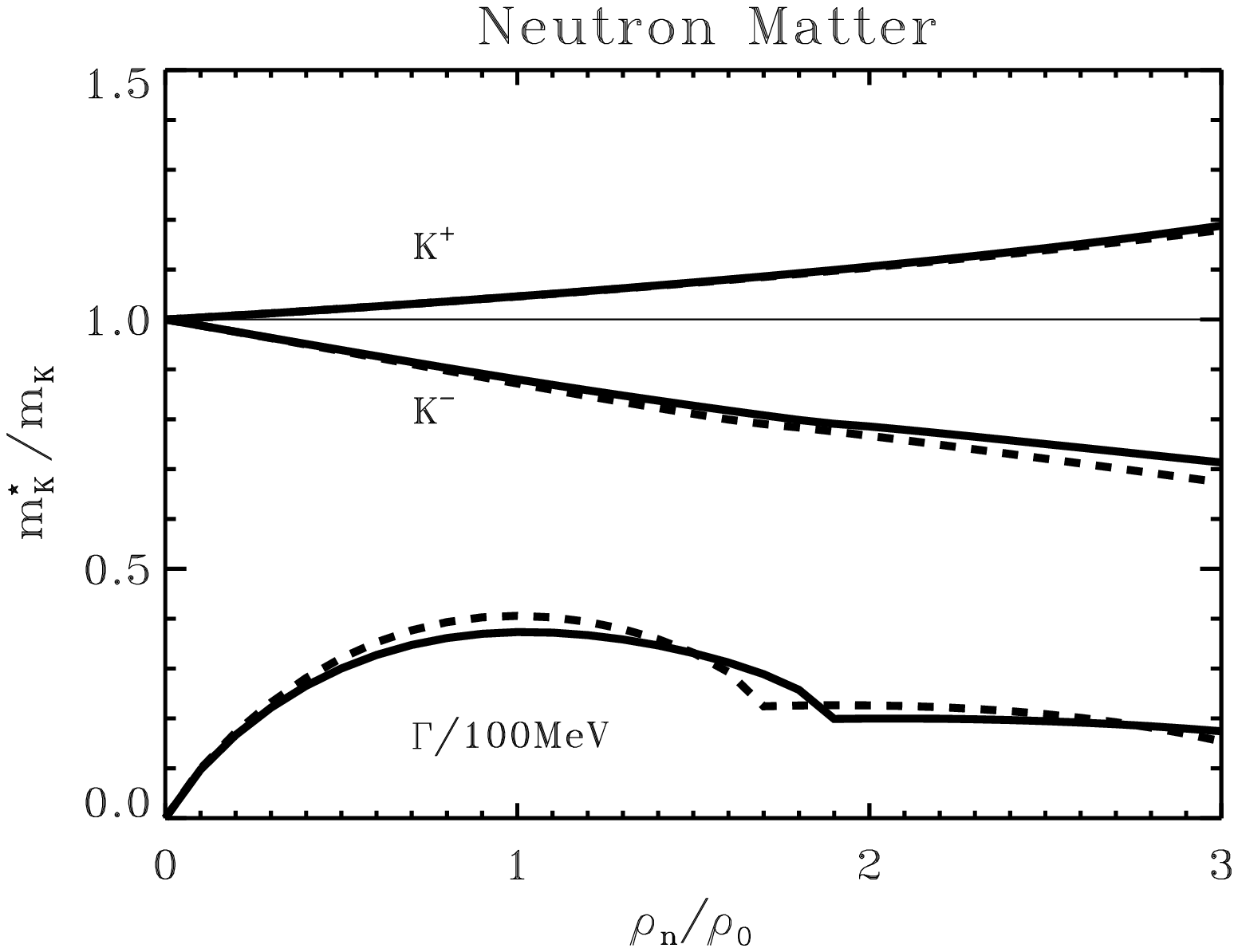}{13.8}
\begin{center}
\vspace*{-.5cm}
\large Figure 4
\end{center}

\newpage
\vspace*{-1cm}
\bild{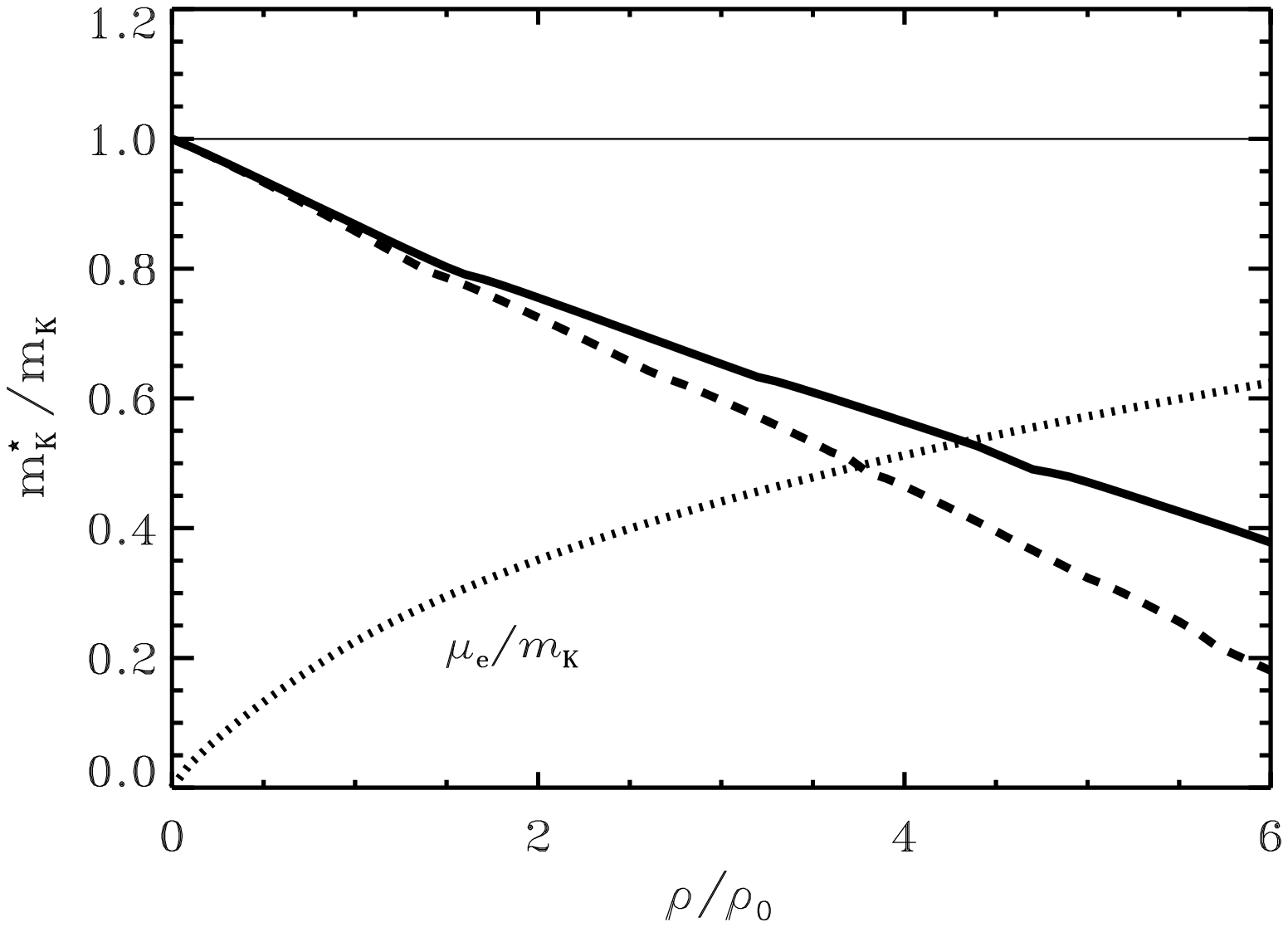}{13.8}
\begin{center}
\vspace*{-.5cm}
\large Figure 5
\end{center}
\end{document}